# Higher order microfibre modes for dielectric particle trapping and propulsion

Aili Maimaiti[1,2], Viet Giang Truong[1], Marios Sergides[1], Ivan Gusachenko[1]

& Síle Nic Chormaic*[1]

[1]Light-Matter Interactions Unit, OIST Graduate University, Onna, Okinawa 904-0495, Japan,
[2]Physics Department, University College Cork, Cork, Ireland

**Optical manipulation in the vicinity of optical micro- and nanofibres has shown potential across several fields in recent years, including microparticle control, and cold atom probing and trapping. To date, most work has focussed on the propagation of the fundamental mode through the fibre. However, along the maximum mode intensity axis, higher order modes have a longer evanescent field extension and larger field amplitude at the fibre waist compared to the fundamental mode, opening up new possibilities for optical manipulation and particle trapping. We demonstrate a microfibre/optical tweezers compact system for trapping and propelling dielectric particles based on the excitation of the first group of higher order modes at the fibre waist. Speed enhancement of polystyrene particle propulsion was observed for the higher order modes compared to the fundamental mode for particles ranging from 1 μm to 5 μm in diameter. The optical propelling velocity of a single, 3 μm polystyrene particle was found to be 8 times faster under the higher order mode than the fundamental mode field for a waist power of 25 mW. Experimental data are supported by theoretical calculations. This work can be extended to trapping and manipulation of laser-cooled atoms with potential for quantum networks.**

## Introduction

Optical trapping and manipulation is an efficient and precise method for controlling and delivering microscopic objects to a desired position, i.e. it can provide stabilizing, accelerating or directional control of the motion of micro- or nanoparticles[1]. The purpose of using this method is to apply the high local confinement of an optical trapping field to specimens of biological interest, such as the manipulation and force measurements of DNA- and RNA-based motors involved in transcription and translation processes[2,3]. Applications include the investigation of cell cytometry[4], artificial fertilization of mammalian cells, and microsurgery[5]. Since Ashkin[6] first introduced optical trapping using a highly focussed laser beam through a high numerical aperture microscope objective, much work has been done based on the so-called optical tweezers[5,7]. Along with that technique, substantial research on optical trapping using evanescent fields in the vicinity of optical waveguides, or at the interface of substrates, has also been undertaken[8–11].

The evanescent field of a waveguide can create strong optical forces, which yield a larger trapping volume at its surface compared to the small waist of a focussed beam in an optical tweezers and this phenomenon overcomes the limitations imposed by the Rayleigh range. Such optical waveguides can provide long range controllable transport of micro/nano-objects[10–12]. Among all waveguides, the optical micro/nanofibre (MNF) is fast becoming a common tool for cold atom and particle manipulation experiments[13] as it can be fabricated in standard laboratory conditions with relative ease and is readily integrated into other fibre optic-based systems.

Until recently, experimental research using MNFs was mostly based on the fundamental guided mode (FM)[13–17]. MNFs initially attracted much interest in the field of cold atoms due to their potential for efficiently coupling light from the trapped atoms to the fibre[18–20]. However, though there have been several theoretical proposals that considered higher order modes (HOMs) for cold atom trapping and detection,[21–23] there was little experimental progress due to the difficulty in fabricating MNFs for efficient HOM propagation. Recently, due to the advances made in higher order mode MNFs fabrication[24–26] preliminary experimental studies on the interactions between atoms and HOMs have been reported[27]. All of this work exploits the longer evanescent field extension from the fibre surface, the three dimensional (3D) arrangement of the field, and the larger cut-off waist diameter of the fibre compared to that needed for fundamental mode propagation. 3D microtrap arrays can be generated when one considers the interference between FM and HOMs co-propagating through an MNF[22,23]. Applications of such 3D potential geometries, in combination with light from cold atoms coupled to the fibre modes, provide a promising method to realise the retrieval and storage of light with atomic ensembles near MNF surfaces.

A 3D HOM potential lattice around a microfibre can also be used to investigate particle manipulation. This all-optical manipulation method offers rapid control of trapped microparticle sites around a microfibre through the generation of radially, azimuthally and hybrid polarised HOM light fields. This provides flexible options for practical applications of particle sorting, control, and the investigation of the internal structure of large complex cells in an aqueous environment. Moreover, the increased interaction of light in HOMs with particles reduces the challenges of fibre handling and increases the lifetime of the fibre since one can work with a micron-sized fibre rather than the more fragile nanofibre. Using HOMs for particle or atom trapping in a multimode waveguide[28] or a hollow core photonic crystal fibre[29] has been studied. The main difficulty in performing these experiments is maintaining the polarisation while HOMs propagate in the fibre. Controlling the loss of HOMs due to the presence of complex trapped objects on the fibre surface can also be challenging. The success of these experiments is, therefore, strongly dependent upon understanding the interaction between each individual particle and the HOM's evanescent fields. However, to our knowledge, the study of the interaction between HOM evanescent fields around a microfibre with individual particles has not previously been demonstrated.

In this paper, we take advantage of the stated benefits of HOMs in MNFs and investigate particle propulsion induced by the evanescent field of HOMs propagating in a microfibre. We used a spatial light modulator (SLM) to produce a doughnut-shaped, first order, Laguerre-Gaussian ($LG_{01}$) beam in free-space. When coupled into suitable optical fibre, $LG_{01}$ excites the $LP_{11}$ approximate, first higher order fibre mode and, at the waist region, the true fibre modes, $TE_{01}$, $TM_{01}$ and $HE_{21e,o}$, are generated[24]. Incorporating the microfibre into an optical tweezers[30] allowed us to control the number of particles in the fibre trap very precisely and study the interaction of one or more particles with the evanescent field at the fibre waist. Propulsion speeds of single polystyrene beads with diameters from 1 μm to 5 μm were compared for both FM and HOM propagation. A speed dependence on particle size was also observed and will be discussed here.

## Theory

**Maxwell's stress tensor and force calculation**

The electromagnetic field distribution can be described using Maxwell's equations. The electric field expression can be determined from equation

$$\nabla \times (\nabla \times \boldsymbol{E}) - k_0^2 n^2 \boldsymbol{E} = 0. \tag{1}$$

Here, $\boldsymbol{E}$ is the electric field, $k_0$ is the wave vector and $n$ is the refractive index of the material (for the case studied here we take $n = 1.45591$ for silica fibre). The interaction between the surface evanescent field of an optical micro/nanofibre and a dielectric micro-object results in three different optical forces acting on the object. The trapping force, also called the gradient force, $\boldsymbol{F}_g$, is induced due to the temporary polarisation of a dielectric particle in a non-uniform field. The other forces are the result of radiation pressure[6]. Namely the scattering, $\boldsymbol{F}_s$, and absorption, $\boldsymbol{F}_a$, forces are responsible for the movement of particles along the optical axis of the fibre. The total force can be described as electromagnetic stress using the Maxwell stress tensor. As particles are transported over a timescale much longer than the optical period of the incident fields coupled into the fibre, the time-independent Maxwell stress-tensor $\langle \boldsymbol{T}_M \rangle$ can be used to calculate the forces. It is given by[31]

$$\langle \boldsymbol{T}_M \rangle = \boldsymbol{D}\boldsymbol{E}^* + \boldsymbol{H}\boldsymbol{B}^* - \frac{1}{2}(\boldsymbol{D} \cdot \boldsymbol{E}^* + \boldsymbol{H} \cdot \boldsymbol{B}^*)\boldsymbol{I}, \tag{2}$$

where $\boldsymbol{E}$, $\boldsymbol{D}$, $\boldsymbol{H}$ and $\boldsymbol{B}$ denote the electric field, the electric displacement, the magnetic field and magnetic flux, respectively. $\boldsymbol{I}$ is the isotropic tensor, $\boldsymbol{D}^*$ and $\boldsymbol{E}^*$ are the complex conjugates.

By applying the relations $\boldsymbol{D} = \varepsilon_r \varepsilon_0 \boldsymbol{E}$ and $\boldsymbol{B} = \mu_r \mu_0 \boldsymbol{H}$, equation (2) can be written

$$T_{ij} = \varepsilon_r \varepsilon_0 E_i E_j^* + \mu_r \mu_0 H_i H_j^* - \frac{1}{2}(\varepsilon_r \varepsilon_0 E_k E_k^* + \mu_r \mu_0 H_k H_k^*), \tag{3}$$

where $E_i$ and $H_i$ are $i$-th components of the electromagnetic field. $\mu_0, \mu_r, \varepsilon_0, \varepsilon_r$ are vacuum and relative permeability and permittivity, respectively.

The total electromagnetic force acting on the particle can be calculated from the integration of the Maxwell stress tensor over the surface of the particle,

$$\boldsymbol{F} = \oint_S (\langle \boldsymbol{T}_M \rangle \cdot \boldsymbol{n}_s) \, dS, \tag{4}$$

where $\boldsymbol{n}_s$ is a normal vector pointing in the outward direction from the surface, $S$.

Here, we consider a system which consists of a particle (with refractive index $\bar{n}_1$) in the evanescent field of a micro/nanofibre in a medium (of refractive index $n_2$). For a non-absorptive particle (i.e. $\bar{n}_1 = \sqrt{\varepsilon_r}$ is real), the force acting on the particle can be expressed as the scattering force ($\boldsymbol{F}_s$) along the fibre axis and the gradient force ($\boldsymbol{F}_g$) perpendicular to it.[32,33]

$$\boldsymbol{F} = \boldsymbol{F}_g + \boldsymbol{F}_s. \tag{5}$$

Here

$$F_g = \frac{1}{4}\text{Re}(\alpha)\nabla|\mathcal{E}|^2, \qquad (6)$$

and $\alpha$ is the electric polarizability of the particle given by $\alpha = 6\pi i \varepsilon_0 a_1^{Mie}/n_2 k_0^3$, $a_l^{Mie}$ being the conventional Mie coefficient[34], $\mathcal{E}$ is the complex amplitude of the electric field.

For nonabsorptive particles $\alpha$ can be approximated as

$$\text{Re}(\alpha) \simeq 4\pi\varepsilon_0 n_2^2 a^3 \frac{\bar{n}_1^2 - n_2^2}{\bar{n}_1^2 + 2n_2^2}, \qquad (7)$$

$$\text{Im}(\alpha) \simeq \frac{8\pi\varepsilon_0}{3} n_2^5 k_0^3 a^6 \left(\frac{\bar{n}_1^2 - n_2^2}{\bar{n}_1^2 + 2n_2^2}\right)^2, \qquad (8)$$

where $a$ is the particle's radius. We assume that the spin component of the field contributing to the force is negligible; therefore, the scattering force can be described as

$$F_s = \sigma_{ext} \frac{n_2}{c} S, \qquad (9)$$

where $S$ is the Poynting vector and $\sigma_{ext}$ is the extinction cross-section of the Rayleigh particle given by[33]

$$\sigma_{ext} = \frac{k_0}{\varepsilon_0 n_2} \text{Im}(\alpha) \simeq \frac{8\pi}{3} n_2^4 k_0^4 a^6 \left(\frac{\bar{n}_1^2 - n_2^2}{\bar{n}_1^2 + 2n_2^2}\right)^2. \qquad (10)$$

**Microfluidic effect on particles near a microfibre surface**

Since the particles move close to the surface of a microfibre, the standard bulk Stokes' drag coefficient, $F = 6\pi\mu a v$, where $\mu$ is the viscosity and $v$ is the particle velocity, cannot be used. Two major correction factors have been used in the literature depending on the distance between the trapped particle and the host surface[35,36]. Assuming $h$ is the distance from the centre of the particle to the fibre surface, Goldman *et al.*[37] used the Faxen correction factor for $h > 1.04\ a$ and the deviation was less than 10% compared to experiment. In this case, the force is given by

$$F = 6\pi\mu a v \left(1 - \frac{9}{16}\frac{a}{h} + \frac{1}{8}\left(\frac{a}{h}\right)^3 - \frac{45}{256}\left(\frac{a}{h}\right)^4 - \frac{1}{16}\left(\frac{a}{h}\right)^5\right)^{-1}. \qquad (11)$$

Alternatively, when $h < 1.04\ a$, the lubrication values of this correction are used, as shown by Krishnan and Leighton[38] such that

$$F = 6\pi\mu a v \left(\frac{8}{15}\ln\left(\frac{h-a}{a}\right) - 0.9588\right). \qquad (12)$$

**Numerical techniques**

A finite element method (FEM) is used to calculate the E-field distribution along the microfibre (equation 1). The wavelength of the propagating light, $\lambda$, was set to 1064 nm. At this wavelength the refractive index of the fibre and water are $n = 1.46$ and $n_2 = 1.33$, respectively. To simplify our calculations, we assumed that the power coupled into each individual component of the HOM set was identical. For all calculations, the total propagating power of the HOM was normalized to 25 mW.

Figure 1 shows the scattering force, $F_s$ (parallel to the fibre surface), and the gradient force, $F_g$ (perpendicular to the fibre surface), acting on the particle. Light propagates from left to right.

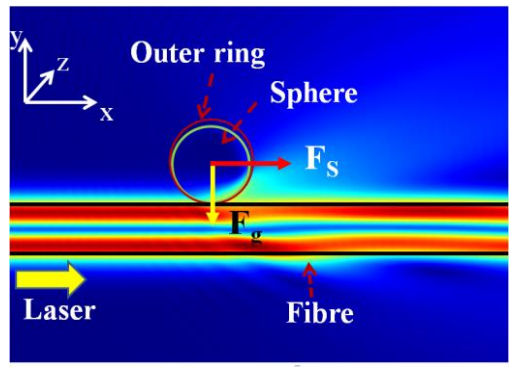

**Figure 1**. **Gradient and scattering ($F_g$, $F_s$) forces on particles in the evanescent field of a microfibre.**

The inner sphere represents the physical polystyrene beads with refractive index $\bar{n}_1 = 1.57$, positioned 20 nm away from the fibre surface. The outer ring represents a circular surface of the momentum transfer region, which is the integration area of the Maxwell stress tensor on the particle[36]. The momentum transfer that causes the particle to be trapped and propelled is equal to the momentum difference between the surrounding medium ($n_2 = 1.33$) and the particle. By changing this outer ring region from 1.05 up to twice the particle size, the total force calculated from this momentum transfer area only varies from 0.3% to 1.5%. The Maxwell stress tensor integration area was therefore set to be 1.1 times the particle size in all our experimentally relevant calculations.

## Results

### Distribution of $LP_{01}$ and $LP_{11}$ modes in a microfibre

According to the weakly guiding approximation[39], modes of a standard fibre are approximated as linearly polarised modes or $LP_{lm}$ modes (where the subscripts l and m indicate the radial and azimuthal order, respectively). However, when the fibre is tapered down and immersed in water, the $LP_{lm}$ modes split in to the $TE_{lm}$, $TM_{lm}$, $HE_{lm}$ and $EH_{lm}$ modes. Here, we consider the $LP_{01}$ or fundamental mode (FM) and the $LP_{11}$ mode group, i.e. the first order higher order modes. The first four HOM profiles and E-field distribution of a 2 μm fibre are shown in figure 2.

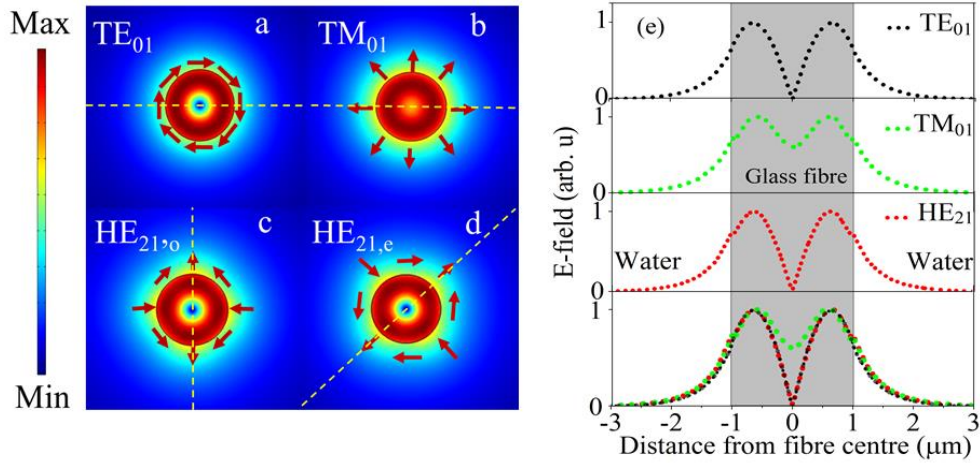

**Figure 2. Mode profiles of the first four true HOMs in 2 μm fibre.** (a-d) Numerical mode intensity profiles of the first four true HOMs. Dashed lines indicate the orientation of the corresponding maximum intensity profile. (e) E-field distributions of each of the individual HOMs (top three distributions), along with a plot showing the distributions superimposed on top of each other (bottom) to show their relative intensities (corresponding to the field along the dashed line).

The calculated decay length for each of the individual HOMs for this fibre shows that all HOMs have approximately same decay length of 0.52 μm (the decay lengths of $TE_{01}$, $TM_{01}$ and $HE_{21}$ are 0.503 μm, 0.52 μm and 0.525 μm respectively) while the FM has a decay length of 0.33 μm.

Figures 3(a) and (b) present the numerical beam profiles and electric field polarisation direction, while figures 3 (c) and (d) show the experimental beam profiles of the fundamental $LP_{01}$ and the higher order $LP_{11}$ mode set, respectively. The $LP_{11}$ mode set in figure 3(b) represents the total E-field distribution of the $HE_{21,o}$, $HE_{21,e}$, $TE_{01}$ and $TM_{01}$ modes. As can be seen from the individual mode profiles (Figure 2), each of the HOMs is different. We chose a specific mode combination as a representative field, consisting of equal portions of each mode. This was used as a model to better explain the phenomena at play in our experiments. Figure 3(e) clearly shows that, when one considers the intensity distribution along the horizontal direction, the total evanescent field of the HOMs have larger amplitudes and longer evanescent field extensions when compared to the FM at a given distance from the surface of a 2 μm fibre. The field intensity, which is calculated from the E-field for a given fibre diameter, also indicates that, if the waist power is the same for the HOMs and the FM, the total evanescent portion of the HOMs is three times larger than that of the FM.

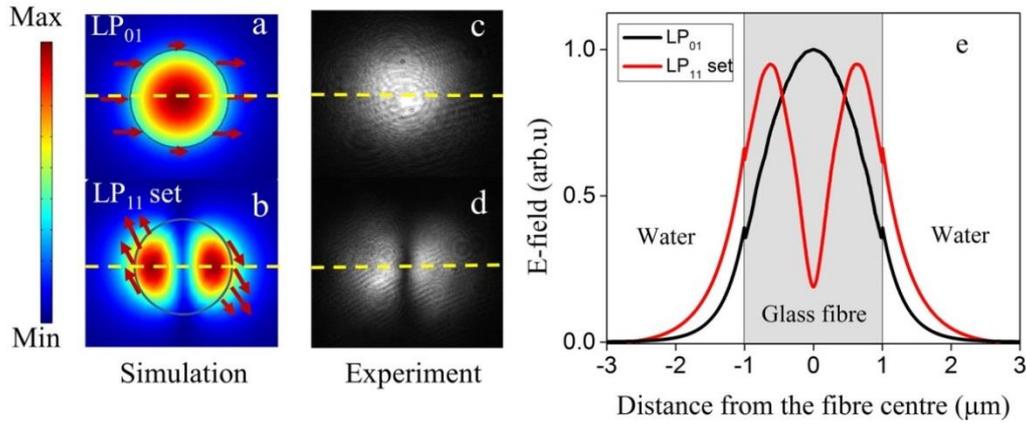

**Figure 3. Field distribution of $LP_{01}$ and the $LP_{11}$ mode set.** (a, b) Numerical mode profiles of the $LP_{01}$ and total E-field distribution of four modes ($TE_{01}$ and $TM_{01}$, $HE_{21,e}$, $HE_{21,o}$) in the $LP_{11}$ set, (c, d) experimental beam profiles of the fundamental $LP_{01}$ and the higher order $LP_{11}$ modes, respectively. The horizontal dashed lines represent the orientation of maximum intensity (e) E-field distribution of the $LP_{01}$ and $LP_{11}$ mode set of a 2 μm microfibre along the direction indicated by the dashed line in (a) and (b).

### Optical forces acting on particles in the evanescent field of $LP_{01}$ and $LP_{11}$ modes

As mentioned in the Theory section, there are two different optical forces acting on the dielectric particles near the microfibre: the gradient force, and the scattering force.

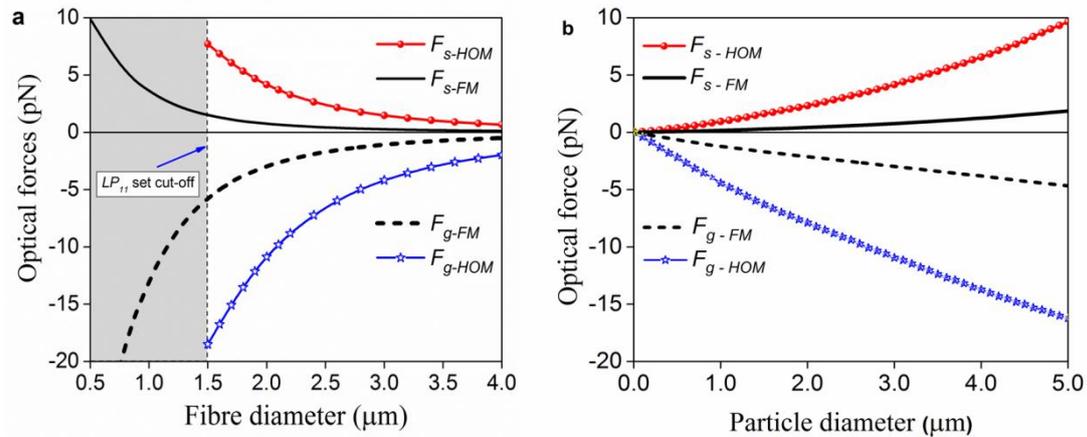

**Figure 4. Calculated optical forces acting on a particle** (a) Magnitude of gradient and scattering ($F_{g\text{-}HOM}$, $F_{s\text{-}HOM}$) forces for the $LP_{11}$ mode set and ($F_{g\text{-}FM}$, $F_{s\text{-}FM}$) forces for $LP_{01}$ mode, respectively, on a 3.0 μm polystyrene particle. Scattering and gradient forces of HOMs are considerably stronger than those of the FM at the HOM cut-off or larger fibre diameter. (b) Particle size dependence of optical forces for the 2 μm fibre diameter. Larger particles experience stronger optical forces. The waist power was 25 mW.

Figure 4(b) presents the calculated scattering ($F_{s\text{-}HOM}$, $F_{s\text{-}FM}$) and gradient ($F_{g\text{-}HOM}$, $F_{g\text{-}FM}$) forces acting on a single particle touching a microfibre in water for both the HOM and FM cases, respectively. Negative values of the gradient force, $F_g$, indicate that the particles are attracted to the fibre surface. A positive scattering force, $F_s$, designates particle propulsion from left to the right along the fibre. It is clearly seen that both absolute gradient and scatter-ing forces decrease with fibre size for all cases. By comparing the forces from figure 4(a), it is found that both forces have greater magnitude in the HOM case compared to the FM case for fibre diameters greater than the $LP_{11}$ cut-off value. This implies a great enhancement of particle trapping stability and propulsion speed in the presence of the HOM evanescent field. Figure 4 (b) shows that both the scattering and gradient forces increase with increased particle size for a given fibre diameter. However, this particle size dependence is more remarkable for HOMs than for the FM.

**Speed of microparticles near a microfibre surface**

Along with the theoretical analysis, an experimental setup was built as in figure 5. A 1064 nm laser was coupled into the microfibre which was immersed in water dispersion. A polystyrene particle was trapped with an optical tweezers and approached to the microfibre. When the tweezers was turned off, the particle was propelled along the fibre. (See Methods and Supplementary Information). To obtain the average particle speed, the position of each particle was recorded for a distance of 75 μm along the microfibre. Each measurement was repeated at least three times.

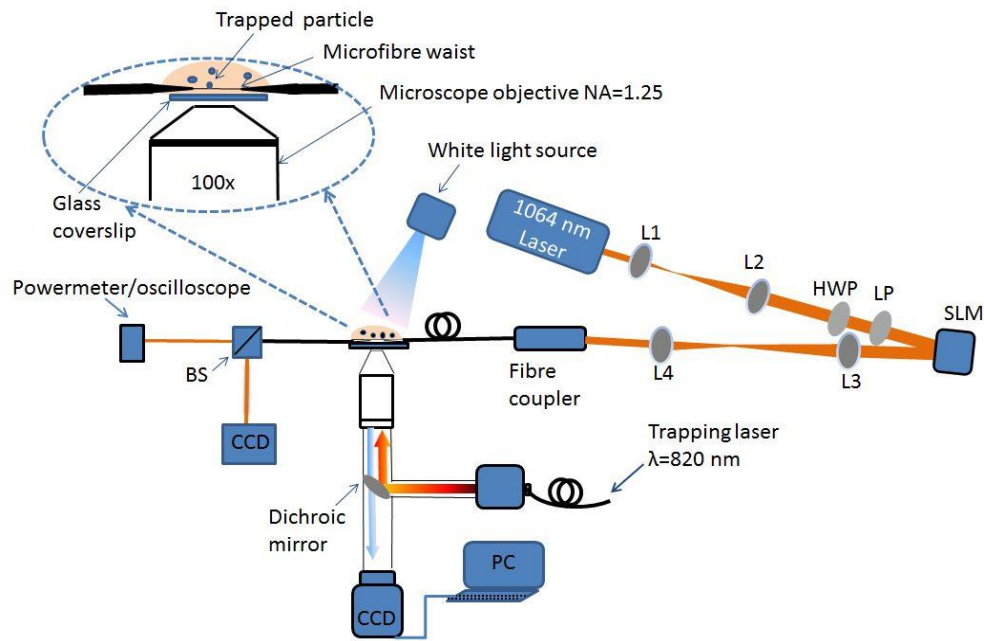

**Figure 5. Experimental setup for particle propulsion.** L1, L2, L3, L4 are lenses, LP and HWP are a linear polariser and a half wave plate, respectively, BS is a beam splitter, CCD is a camera and PC is a computer. The red lines represent the free beam path. The inset shows the magnified image of the interaction region.

Figures 6(a) and (b) show the consecutive images of a propelled 3 μm polystyrene particle captured by the CCD camera under both FM and HOMs conditions. In this time interval (1.2 s), when the waist power was at 25 mW, the particle underwent very slow propulsion under FM propagation while the speed of the particle was greatly increased under HOM propagation. The frame rate of the camera used was not fast enough to track propelled particles at high speed. This caused the particles to appear elongated along the propulsion axis (see Figure 6(b))

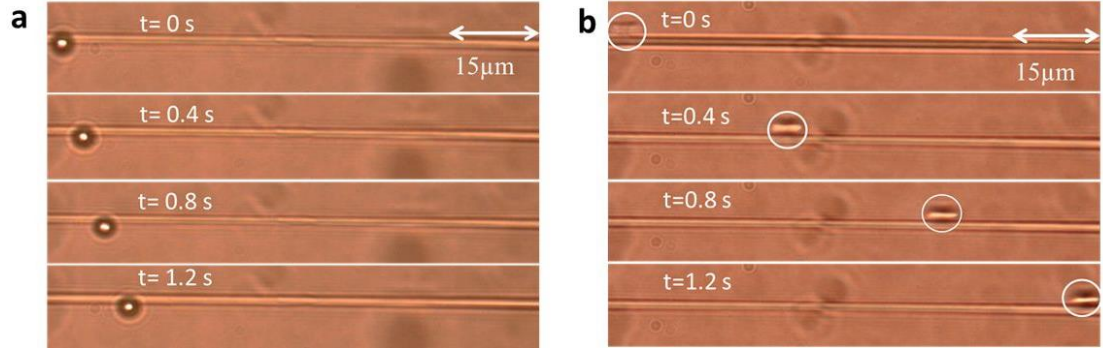

**Figure 6. Micrograph of 3 μm polystyrene particle propulsion under** (a) FM and (b) HOMs propagation. In both cases the waist power was 25 mW.Table 1 gives the speed of a 3 μm particle at a fixed waist power of 25 mW obtained experimentally and by three different theoretical methods. The experimental results confirm that, under identical powers at the microfibre waist, the particle speed is approximately 8 times faster under $LP_{11}$ mode set propagation than in the $LP_{01}$ case. Such an increase in speed is indicative of the stronger evanescent field and larger field extension of the HOMs.

It is apparent from the values in Table 1 that the bulk Stokes drag coefficient cannot be used to describe the experimental results. In order to better understand the phenomenon of microfluidic approach on particles moving along a microfibre surface, two correction factors were used to calculate the particle velocities[37,38]. It is clearly shown in Table 1 that, for the FM case, the Krishnan model is in reasonable agreement with the experiment for low particle speeds when the particle is close to the fibre surface ($h = 1.52$ μm). On the other hand, under the HOM's evanescent field, the speed of the particle is relatively fast compared to the FM case. This leads to strong hydrodynamic interactions that may push the particle to float at a considerable distance from the microfibre surface due to the hydrodynamic lift force. Here, the Krishnan approximation does not hold and the experimentally obtained speed seems to be closer to the value approximated in the Faxen region for $h = 1.56$ μm.

Figure 7 shows the change in the speed of a single 3 μm polystyrene sphere under the evanescent field of both FM and HOMs with varying power as measured from the recorded video files. The agreement between the experimental results and the Krishnan and Faxen corrections is evident in the FM case. However, the particle speeds for the HOMs is slightly higher than those obtained by the Faxen approximation for waist powers less than 25 mW. We assume that the increase in speed due to the increase in waist power causes the particle to move further away from the fibre surface. As this distance increases, the surface and/or hydrodynamic interaction between the particle and the fibre walls weakens allowing the particle to reach a higher speed.

With the FM waist power less than 20 mW, particles tend to move in discrete jumps or appear at stable positions along the fibre surface. This is probably due to surface roughness effects that can easily dominate over optical forces at low powers.

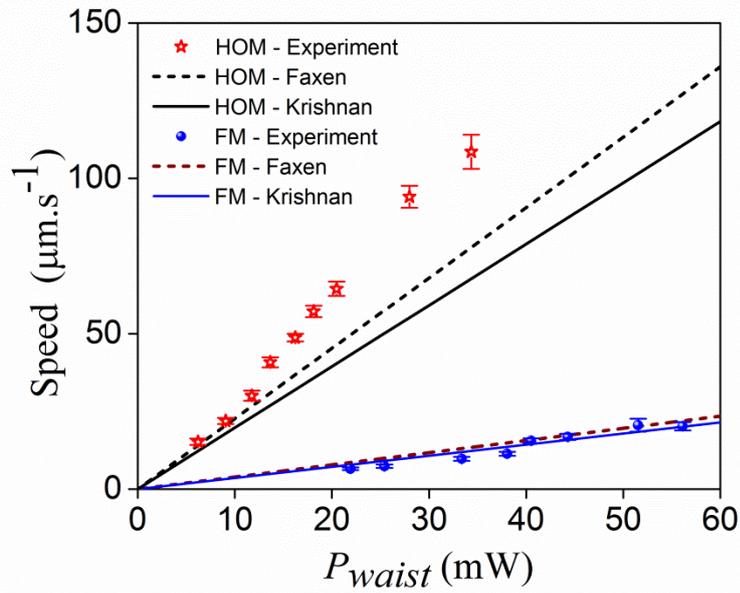

**Figure 7. Measured speed of a 3 μm polystyrene particle under the evanescent field of the FM (blue circles) and HOM (red stars) as a function of waist power, $P_{waist}$.** The speed is significantly higher for HOM propagation. The filled circles and unfilled stars represent the experimental data and the lines are the theoretical plots.

**Effect of particle size on speed of a particle in the microfibre evanescent field**

In order to better understand the power/speed relationship described above, and characterize the speed changes with different size particles under HOM and FM propagation, polystyrene particles with diameters of 1 μm, 2 μm, 3 μm and 5 μm were experimentally examined under the same conditions. Figure 8 shows the speed dependence on the particle size for both HOM and FM evanescent field.

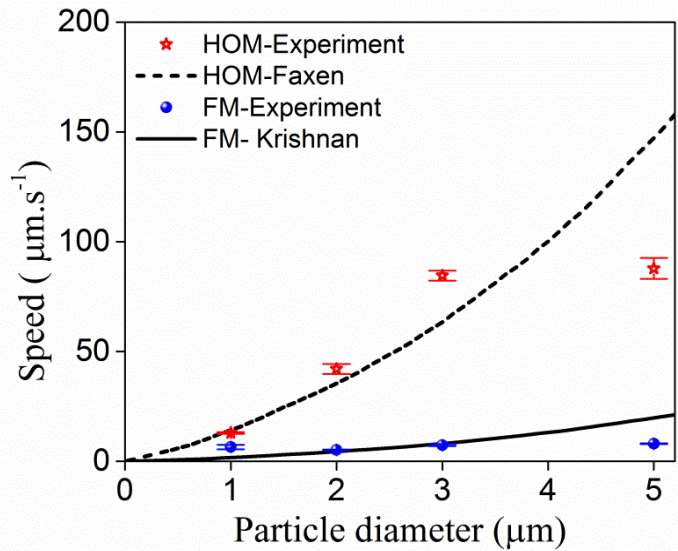

**Figure 8. Measured speed dependence on particle size for HOM (blue circles) and FM (red stars) propagation.** Here, the waist power was set to 25 mW for all cases. The lines are theoretical curves.

It is evident that, for HOMs, an increase in particle size is consistently matched with an increase in speed. This size dependence of particle speed has also been observed for the fundamental mode when the fibre size is comparable to the guided wavelength[14]. It is well known that larger particles have larger scattering cross-sections compared to smaller particles. Consequently, they have a greater interaction with a given evanescent field, causing them to have faster speeds than smaller particles[14]. It is interesting to note that the experimentally observed increase in speed with larger particle size is not linear. In figure 8, the speed of a 5 μm particle was smaller compared to the theoretical estimates for both HOMs and FM. This could be due to the fact that, as the particle diameter increases, the surface interaction between the particles and the fibre surface also increases leading to lower propulsion speeds along the fibre.

## Discussion

Fabrication of high-quality tapered optical fibres for FM propagation is now a routine practice, with laboratories reporting up to 99.95% light transmission through the fibre at the structure waist[40]. However, tapered fibres that support HOMs with high transmission have only recently been experimentally realized[24–26]. This work presents the first experimental results using highly transmitted HOMs for particle propulsion. The experimental setup combines three subsystems which allowed us to overcome the challenge of generating and obtaining high quality HOMs through an adiabatic microfibre while minimising system interference to study single particle dynamics.

A complete microfluidic system would include the particle's angular rotation, the change of intensity ratio between the HOM components while propagating along the waist region of the microfibre, the surface roughness and the optical torque. It is known that the FM evanescent power of a 2 μm fibre diameter is small and particle speed is relatively slow. Particles that propel with a low speed are closer to the fibre surface due to weak hydrodynamic interactions. In this case, the surface roughness may cause the particle motion to suffer from frictional forces[38]. The lubrication correction factor, therefore, seems to fit with experiment.

The ratio of the HOMs to the FM velocities, as calculated, is found to be around 5.7 using either method while in experiments this ratio is around 8. This discrepancy can be explained in terms of hydrodynamic interactions. When a particle is moving along a fibre, there are many parameters such as translational speed, particle rotation, surface roughness, etc. which can contribute to the separation distance between the particle and fibre surface; this in turn influences the particle speed. Moreover, when a particle is propelled at high speed, the hydrodynamic lift force can oppose the attractive optical gradient force. This allows the particle to have a larger separation distance from the fibre[35,37,38,41]. Therefore, under high HOM power (>20 mW), the particle speed may be faster than that calculated in simulations (see Figure 7).

Furthermore, changes in the ratio between the HOM components while propagating along the fibre will change the amplitude of the $LP_{11}$ set due to the mode interference. In an ideal situation the proportion of each of the modes in the $LP_{11}$ mode set in the fibre is assumed to be equally distributed. In reality, the proportion of each mode in the fibre may change, leading to a change in the total optical force. In the simulations where $F_s$ and $F_g$ were calculated it was assumed that this ratio remained constant; this may be introducing further inconsistency with the experimental results. The interaction of each of the individual modes in the HOM set with the particle is beyond the scope of this work.

In conclusion, we have successfully demonstrated particle propulsion under the evanescent field of the HOM of a microfibre and compared it with the FM case. Both theoretical analysis and experimental observations showed a higher speed for microparticles under the influence of the evanescent field of HOMs. We also used different sizes of polystyrene beads to address speed dependence on particle size. The hydrodynamic interaction between particles and fibre surface has been studied and discussed. Two correction factors of the hydrodynamic interaction effect have been used to predict the particle speed for the HOM and FM cases. Apart from these, a number of other parameters need to be addressed for a complete understanding of the experiment such as optical torque and surface roughness. An important experimental feature of this work consists of the integration of a MNF within a standard optical tweezers. Specifically, this facilitated the study of individual particle propulsion. As this experimental framework provides exact control of the number of particles that interact with the fibre, in future we plan on studying optical binding and collective propulsion [17] on interaction with HOMs.

The HOMs have the advantage of possessing a strong evanescent field, thus increasing the sensitivity of interaction between the field and the particles at robust microfibre diameters compared to more fragile nanofibres. The larger evanescent field decay and the field distribution of HOMs can be utilised for particle and refractive index sensing where high sensitivity is required[42,43]. Additionally, this work can be developed beyond colloidal particles to study novel cold atom traps.

## Methods

### Generation of higher order modes

To excite the HOMs in the fibre, a beam with a doughnut-shaped intensity cross-section was created in free space. Linearly polarised 1064 nm laser light (Ventus 1064 3W) was launched onto the display of an SLM to create a first-order Laguerre-Gaussian beam ($LG_{01}$) at the far field[44]. Two-mode fibre (Thorlabs, SM1250G80) operating at 1064 nm wavelength with a cladding diameter of 80 μm was chosen for this experiment. The fibre supports the fundamental $LP_{01}$ and the $LP_{11}$ family of higher order modes. By coupling the $LG_{01}$ beam to this fibre, a two-lobed pattern corresponding to the $LP_{11}$ mode can be obtained at the fibre output. By switching the vortex phase grating on the SLM on and off, the propagating mode in the fibre can be easily switched between LP01 and $LP_{11}$ with high purity. By choosing an appropriate objective lens, 45% coupling efficiency to the $LP_{11}$ mode was achieved. A charge-coupled device (CCD) camera was used to image the quality of the beam profile. Figure 4 shows the schematic of the experimental setup.

### Preparation of higher-mode tapered fibre

Tapered fibres were made on a fibre pulling rig system that uses the flame brush technique, with an oxygen/hydrogen gas source[26]. The taper shapes may be custom designed with a Matlab code, based on analysis of the required shape[26,45]. Tapered fibres which support HOMs have been experimentally realised for the above mentioned 80 μm cladding diameters. Based on recent theoretical work[24,45], and using improved experimental methods for fabricating adiabatic HOM tapers, an 80 μm fibre was tapered down to a diameter of 2 μm with two consecutive taper angles of 0.6 mrad and 1 mrad[26]. These taper angles were used to reduce loss caused by undesired coupling of the modes to higher order symmetric cladding modes, and to avoid having an excessively long taper. During the tapering process, the output transmission was measured and recorded simultaneously by using a photodiode connected to an oscilloscope. By slowly tapering the fibre, typical transmission of 80% of the $LP_{11}$ mode was ob-tained at the fibre output. By switching back to the FM mode via SLM modulation, we observed that the microfibre still transmitted 95% of the fundamental mode.

According to the commonly accepted analysis used for determining the power propagating through the waist of a homogeneous tapered fibre, the total power loss of a uniformly tapered fibre originates only from the transition regions of the microfibre which are symmetrically located on either side of the microfibre waist. The total power at the microfibre waist can therefore be estimated as equal to the square root of the product of the input and output powers.

**Integrating the optical tweezers with the tapered fibre**

To introduce a single particle to the microfibre surface[30], and minimise system interference from unwanted particles, an optical tweezers was built. Trapping was done via an inverted 100x microscope objective (NA = 1.25). A thin glass coverslip was positioned at the focal plane of the objective to support the liquid sample and trap particles. A Ti: Sapphire laser source operating at 820 nm was used to provide a stiff trap.

The prepared microfibre was fixed to a mount on a 3D translation stage to allow both vertical and horizontal fibre adjustment over the trapping plane (Fig. 5). A CCD camera and a white light source were used to observe the system via the tweezers objective.

Polystyrene particles with diameters of 1 μm, 2 μm, 3 μm, and 5 μm were prepared in deionized water dispersion that prevents particles from sticking on the surface of the fibre causing undesirable power loss. The optical tweezers was used to trap a single particle and convey it to the evanescent field of the microfibre. As soon as the trapped particle is released from the tweezers, the evanescent field of the microfibre propels the particle in the direction of light propagation. By observing these dynamics, the speed of each single particle under both the HOM and FM transmission was investigated (see Supplementary Information).

## Acknowledgments

This work was supported in part by funding from the Okinawa Institute of Science and Technology Graduate University. MS acknowledges the support of JSPS through the Postdoctoral Fellowship for Overseas Researchers scheme. The authors thank M. Frawley, J. Ward, and A. Shen for invaluable discussions and suggestions.

**Author contributions**

S. N. C proposed and supervised the project; A. M and V. G. T equally contributed to the experiments and numerical calculations; all authors participated in data analysis and manuscript writing. All authors reviewed the manuscript.

**Competing financial interests**: The authors declare no competing financial interests.

**Table 1** | **Hydrodynamic velocities of single 3 μm polystyrene beads.** Speed of particles is calculated using different correction factors for the $LP_{01}$ and $LP_{11}$ microfibre modes. The power at the fibre waist regions $P_{waist}$ was fixed at 25 mW.

| Modes | Particle radius $a$ (μm) | Distance from particle centre to fibre surface, $h$ (μm) | Scattering force $F_s$ (pN) | Correction | Velocity (μm/s) |
|---|---|---|---|---|---|
| $LP_{01}$ | 1.5 | 1.52 | 0.75 | Stokes | 29.0 |
|  |  | 1.52 | 0.75 | Krishnan | 8.9 |
|  |  | 1.56 | 0.67 | Faxen | 9.6 |
|  |  |  |  | Experiment | 8.5 |

| LP$_{11}$ | 1.5 | 1.52 | 4.15 | Stokes | 160.5 |
|---|---|---|---|---|---|
| | | 1.52 | 4.15 | Krishnan | 49.2 |
| | | 1.56 | 3.90 | Faxen | 56.0 |
| | | | | Experiment | 72.5 |